\documentclass[12pt,reqno]{amsart}%{article}\nonstopmode% 
\usepackage{amsmath,amssymb,amsfonts,amsthm,eucal,amscd}

\def\boxx#1{\leavevmode\vbox{\hbox to 0pt{\hss\raise1.8ex\vbox 
to 0pt{\vss\hrule\hbox{\vrule\kern.75pt\vbox{\kern.75pt\hbox{\tiny #1}\kern.75pt}\kern.75pt\vrule}\hrule}}}\relax} 

\def\LL#1{\label{#1}\protect\boxx{#1}} 
\def\LL#1{\label{#1}}

\numberwithin{equation}{section}

\newtheorem*{theorem*}{Theorem} 
\newtheorem{theorem}{Theorem} 
\newtheorem{lemma}{Lemma}[section] 
\newtheorem{proposition}[lemma]{Proposition} 
\newtheorem{corollary}[lemma]{Corollary} 
\theoremstyle{definition} 
\newtheorem{definition}[lemma]{Definition} 
\newtheorem{example}[lemma]{Example} 
\newtheorem{notation}[lemma]{Notation} 

\newtheorem{remark}[lemma]{Remark} 
\theoremstyle{remark} 

\begin{document}

%%%%%%%%%%%%%%%%%%%%%%%%%%%%%%%%%%%%%%%%%%%%%%%%%
\long\def\BTHM#1#2{\begin{theorem}\LL{#1}#2\end{theorem}}
\long\def\BDF#1#2{\begin{definition}\LL{#1}#2\end{definition}}
\long\def\BPROP#1#2{\begin{proposition}\LL{#1}#2\end{proposition}}
\long\def\BLEM#1#2{\begin{lemma}\LL{#1}#2\end{lemma}}
\long\def\BNT#1#2{\begin{notation}\LL{#1}#2\end{notation}}
\long\def\BREM#1#2{\begin{remark}\LL{#1}#2\end{remark}}
\long\def\BEX#1#2{\begin{example}\LL{#1}#2\end{example}}
\long\def\BCOR#1#2{\begin{corollary}\LL{#1}#2\end{corollary}}
\long\def\BEQ#1#2{\begin{equation}\LL{#1}#2\end{equation}}
\long\def\BB#1#2#3{\begin{#1}\LL{#2}#3\end{#1}} 
\def\Pr{\begin{proof}}
\def\rP{\end{proof}} 
%%%%%%%%%%%%%%%%%%%%%%%%%%%%%%%%%%%%%%%%%%%%%%%%%

\title[On the theorems of Krein and Sakhnovich] 
{A note on the theorems of M.~G.~Krein\\and L.~A.~Sakhnovich 
on 
continuous analogs\\of orthogonal polynomials on the circle} 
\author[A. Teplyaev]{Alexander Teplyaev}

\address{Department of Mathematics, 
University of Connecticut, Storrs CT 06269 USA}

\email{teplyaev@math.uconn.edu}

\thanks{Research supported by the 
National Science Foundation (grant DMS–0071575).}

\date{\today}

\begin{abstract}
\noindent
Continuous analogs of orthogonal polynomials 
on the circle are solutions of a canonical system 
of differential equations, introduced and studied by 
M.G.Krein and recently generalized to matrix systems by 
L.A.Sakhnovich. 
We prove that the continuous analogs of the adjoint polynomials converge 
in the upper half-plane in the case of $L^2$ coefficients, 
but in general the limit can be defined only up to a constant 
multiple even when 
the coefficients are in $L^p$ for any $p{>}2$, the spectral measure 
is absolutely continuous and the 
Szeg\"{o}--Kolmogorov--Krein condition is satisfied. 
Thus we point out that 
Krein's and Sakhnovich's papers contain an inaccuracy, 
which does not undermine known implications from these results. 
\tableofcontents
\end{abstract} 

\maketitle

\section{Introduction.} 
Orthogonal polynomials on the unit circle have interesting 
features that relate properties of their spectral measure to the 
properties of coefficients of generating recursive formulas 
(see Section~\ref{sPoly} for more details). 
The present paper deals with continuous analogs of such polynomials. 

The one dimensional analogs were introduced by 
M.~G.~Krein in [{K}]. They provide, in a sense, 
a generalization of the Fourier transform from $L^2(\mathbb R)$ 
to $L^2(\mathbb R, \tau)$. Here $\tau$ is 
%an arbitrary, up to a mild constrain (\ref{eee}), 
a Borel spectral measure 
on $\mathbb R$. 
In this generalization of the Fourier transform, 
the usual exponentials $e^{\text{\bfseries\itshape i\/\/} r\lambda}$ 
are replaced with $p(r,\lambda)$, 
the continuous analog of orthogonal polynomials. 
We consider only ``one sided'' situation, that is, 
$r$ is nonnegative and the Fourier transform is from a half-line 
to the whole line (see Section~\ref{sK}). 

Note that the Fourier transform itself is a continuous analog of 
the expansion into the Fourier series, 
insofar as 
$$\{e^{\text{\bfseries\itshape i\/\/} r\lambda}\,|\,
r\in\mathbb R_+,\lambda\in\mathbb R\}$$ 
are analogous to 
$$\{z^n\,|\,n\in\mathbb Z_+,|z|=1\}.$$ 
Similarly, $$\{p(r,\lambda)\,|\,
r\in\mathbb R_+,\lambda\in\mathbb R\}$$ 
 are analogous to $$\{\varphi_n(z)\,|\,n\in\mathbb Z_+,|z|=1\},$$ 
orthonormal polynomials of degree $n$  on the unit circle 
with respect to an arbitrary probability Borel spectral measure $\tau$. 
To add one more analogy, note that $\varphi_n(z)=z^n$ are the 
orthogonal polynomials with the 
normalized Lebesgue measure as the spectral measure. 

In [S1--S5] L.~A.~Sakhnovich defined and studied matrix valued continuous an\-a\-logs of 
orthogonal polynomials on the unit circle, and 
generalized Krein's results for this case 
(see Section~\ref{sSakh}).

The functions $p(r,\lambda)$, together with 
the continuous analog $p^*(r,\lambda)$ of the adjoint polynomials, 
are solutions of a canonical 
system of differential equations (\ref{eK}). 
The spectral measure $\tau$ is 
uniquely determined by these differential equations. 
The Krein differential equations are related to the 
study of the one dimensional continuous 
Schr\"odinger equation 
[{D1,D5,DK2,K}]. Also they can be used to solve an
important factorization problem in the theory of analytic functions 
[{A,DK1,G,Sz,Si}].

As an expository remark, we note that another way to define $p(r,\lambda)$ and $p^*(r,\lambda)$ is by 
the formulas 
$$
p(r,\lambda)=
e^{\text{\bfseries\itshape i\/\/} r\lambda}\Big(1-\int_0^r \Gamma_r(s,0)
e^{-\text{\bfseries\itshape i\/\/} s\lambda}\Big) ds 
$$
$$
p^*(r,\lambda)=1-\int_0^r 
\Gamma_r(0,s) %-corrected by D.
e^{\text{\bfseries\itshape i\/\/} s\lambda} ds. 
$$
Here $\Gamma_r(s,t)=\overline{\Gamma_r(t,s)}$ is the resolvent
 of a positive integral operator $S_r$, that is 
$$
\Gamma_r(s,t) + \int_0^r H(s-u) \Gamma_r(u,t)du= H(s-t),
$$
where $H(t)=\overline{H(-t)}$  and %-corrected by D. 
%-is locally integrable 
$$
S_rf(x)=f(x)+\int_0^r H(x-t) f(t) dt . 
$$
The coefficient $a(r)$ of the equation (\ref{eK}) is 
$a(r)=\Gamma_r(0,r)$. 
Usually, the accelerant $H(t)$ 
is assumed to be continuous to construct the corresponding Krein system
with continuous coefficient $a(r)$. In our work we do not use such a construction, 
but define $p(r,\lambda)$ and 
$p^*(r,\lambda)$ as solutions of Krein's canonical 
system of differential equations (\ref{eK}). 
%Thus we need only local integrability of $a(r)$ to have the 
%existence of $p(r,\lambda)$ and $p^*(r,\lambda)$. 

If $\tau'$ is the density of the absolutely 
continuous component of the spectral measure, then the Szeg\"{o}--Kolmogorov--Krein condition
\BEQ{ei1}{\int_{\mathbb R} \frac{| \log \tau'(\lambda)|} 
{1+\lambda^2} d\lambda < \infty }
is satisfied if and only if 
\BEQ{ei2}{\int_{0} ^\infty |p(r,\lambda)|^2 dr < \infty} 
for $\text{Im}\lambda>0$. Notice that no assumption on the singular part of $\tau$ is made except (\ref{eee}). 

In the center of our discussion is the existence of the limit 
\BEQ{epi}{
\Pi(\lambda) = \lim_{r\to\infty} p^*(r,\lambda),}
where 
$\Pi(\lambda)$ is analytic for $\text{Im}\lambda>0$. 
M.~G.~Krein pointed out in \cite{K} that  if the coefficients are square 
integrable, then the limit (\ref{epi}) converges. 
In Section~\ref{SBocv} 
we prove that this so even in the matrix case,  and therefore  
$\Pi(\lambda) $ is uniquely defined for square 
integrable coefficients. 
Section~\ref{SBocv} also contains other results related to the 
convergence of the limit (\ref{epi}) in the case of the 
the Sakhnovich differential equations. 

An important relation, which follows from 
(\ref{eK}) and was noted by M.~G.~Krein in [{K}], is 
\BEQ{eKK}{|p^*(r,\lambda)|^2-|p(r,\lambda)|^2 = 
2\text{Im}{\lambda}\int_{0}^r |p(s,\lambda)|^2ds.} 
This a particular case of Lagrange identity, which is 
an analog of the 
Christoffel-Darboux 
formula for orthogonal 
polynomials (see, for instance, [At]). 
Thus we must have $$|\Pi(\lambda)|^2= 
2\mbox{Im}\lambda\int_0^\infty |p(r,\lambda)|^2dr$$ 
if the integral converges and the limit \eqref{epi} exists. 

The existence of the limit (\ref{epi}) implies 
the convergence of the integrals (\ref{ei1}) and (\ref{ei2}), 
but the converse is not true in general. 
In Section~\ref{Sce} we prove that there 
are situations when (\ref{ei1}) and (\ref{ei2}) hold, but 
$\Pi(\lambda)$ has to be defined as a limit of a convergent subsequence. 
We show that this situation is not ``pathological'', but 
can occur even if the spectral measure 
$\tau$ is absolutely continuous with positive continuous density 
(Theorem~\ref{thm111}). In another example (Theorem~\ref{thm12}), 
 this happens even though 
$$ 
\lim_{r\to\infty} |p^*(r,\lambda)|^2=|\Pi(\lambda)|^2, 
$$ 
and the coefficients are in $L^p$ for any $p>2$. 
Moreover, the function $\Pi(\lambda)$ can not 
be defined uniquely, but only up to a constant factor of absolute 
value one (up to left multiplication by a unitary matrix 
in the case of the Sakhnovich theorem).

Note that results of Section~\ref{SBocv} apply to 
 the Krein system, since it is a particular case of the Sakhnovich system. 
Two of the three results there are new even for the Krein system. 
At the same time results of Section~\ref{Sce} are stated for 
 the Krein system, but are applicable for the Sakhnovich system as well.

The fundamental paper [{K}] presents a number of 
important 
results, 
%analogous to the theorems on the orthogonal polynomials 
%on the unit circle
though it does not contain proofs due to the type of the journal 
it was published in. 
Later proofs of Krein's results were given 
%, in part, by N. I. Akhiezer, A. M. Rybalko in [{AR,R}]. 
%Complete proof were 
independently by the author in 1990 
(partly published in [{T1}]) 
and 
L.~A.~Sakhnovich in 1998 ([S2--S4]). 
The main subject of [{T1}] 
was to prove that the spectral measure $\tau$ is absolutely continuous 
with probability one if the coefficient $a(r)$ is a random 
function satisfying certain conditions. 

In [{T1}] 
the author noted and rectified an inaccuracy in the statement of 
Krein's theorem, and gave a proof of the corrected main theorem 
(see Section~\ref{sK} for more details). 
Theorems \ref{thm111} and \ref{thm12} 
in Section~\ref{Sce} prove, in particular, that a part of the statement of 
 the Krein theorem in [{K}] needs to be revised. 

In [S1--S5] L.~A.~Sakhnovich defined and studied matrix valued continuous an\-a\-logs of 
orthogonal polynomials on the unit circle, and 
proved matrix generalizations of Krein's results. 
Unfortunately, these works contain 
the same kind of inaccuracy as~[{K}]. 
% (see Remark \ref{remLAS}). 
In Section~\ref{sSakh} we present the corrected statement, and 
 the corrected part of the proof. 

We emphasize that the inaccuracy in the statement of 
 Krein's and Sakhnovich's theorems is not significant, and does not 
undermine known implications from these important results. 
For instance, if (\ref{ei1}) and (\ref{ei2}) hold, then there is the function 
$\Pi(\lambda)$ which is analytic and has no zeros 
for $\text{Im}\lambda>0$, and 
$$
\tau'(\lambda) = \frac{1}{2\pi|\Pi(\lambda)|^2}
$$ 
for Lebesgue almost all $\lambda\in\mathbb R$ 
(there is an analogous matrix version proved by L.~A.~Sakhnovich in 
[{S4}]). 
This result remains unchanged even if the 
limit (\ref{epi}) 
%${\lim\limits_{r\to\infty} p^*(r,\lambda)}$ \smash
diverges, 
and the nonuniqueness of $\Pi(\lambda)$ mentioned above takes place. 

\subsubsection*{Acknowledgments.} The author is deeply grateful 
to I.~A.~Ibragimov for his invaluable guidance during an early part 
of this work, which was completed in St.-Petersburg, former Leningrad, 
State University and Steklov Mathematical Institute (POMI). 
The author 
thankfully acknowledges many insightful remarks and suggestions of S.~A.~Denisov. 
The author is  grateful to M.~I.~Gordin, M.~L.~Lapidus, 
L.~A.~Sakhnovich 
and 
M.~Z.~Solomyak 
for helpful discussions during the preparation of this paper. 

%Sakhnovich, Lev Aronovich*** 

\section{Orthogonal polynomials on the circle.\LL{sPoly}}

If $\{\varphi_n(z)\}_{n=0}^\infty$ are 
%monic 
polynomials of degree $n$, orthonormal on the unit circle 
with respect to a probability Borel measure $\tau$, then there 
exists a  sequence of complex numbers $\{a_n\}_{n=0}^\infty$ 
such that the following recurrent relations hold 
\BEQ{e1}{
\begin{aligned}
\varphi_{n+1}(z)&=(1-|a_n|^2)^{-1/2}\big(z\varphi_n(z)-\bar a_n \varphi_n^*(z)\big)\\
\varphi_{n+1}^*(z)&=(1-|a_n|^2)^{-1/2}\big(\varphi_n^*(z)- a_n z \varphi_n(z)\big)
\end{aligned}}
with initial conditions 
$$
\varphi_0(z)=\varphi_0^*(z)=1.
$$ 
The auxiliary polynomials $\varphi_n^*(z)$ 
are adjoint to the orthogonal polynomials $\varphi_n(z)$ in the sense that 
$\varphi_n^*(z)=\bar c_0 z^{n}+\ldots+ \bar c_j z^{n-j} +\ldots+ \bar c_n$ \,if\, 
$\varphi_n  (z)=     c_0      +\ldots+      c_j z^{j}   +\ldots+ c_n z^n$. 

The so called 
circular (reflection, Shur's) parameters $\{a_n\}_{n=0}^\infty$ 
satisfy 
%\mbox{$|a_n|\leqslant1$}. Moreover, 
\BEQ{e2}{|a_n|<1}
 for all $n$ if and only if the measure $\tau$ is not concentrated in a 
finite number of atoms. 
Conversely, if conditions \eqref{e2} are satisfied, then 
there exists a unique Borel probability measure $\tau$ on the unit circle 
such that polynomials $\{\varphi_n(z)\}_{n=0}^\infty$, defined by \eqref{e1}, 
 are orthonormal with respect to $\tau$. 

The theory of orthogonal polynomials on the circle was developed by 
G.~Szeg\"{o}, N.~I.~Akhiezer, 
L.~Ya.~Geronimus \emph{et al}. ([{A,G,Sz}]). The following 
theorem is a combination of 
 results of G.~Szeg\"{o}, A.~N.~Kolmogorov, M.~G.~Krein 
and L.~Ya.~Geronimus (see [{G,Si}]).

\begin{theorem*} 
The linear span of 
 $\{\varphi_n(z)\}_{n=0}^\infty$ is not dense in $L^2_\tau$ if and only 
if any of the following five equivalent statements hold 
\begin{itemize}
\item[(I)] \BEQ{e3}
{\int_0^{2\pi}\log \tau'(e^{\text{\bfseries\itshape i\/\/}\,\theta}) d\theta 
> -\infty} 
where $\tau'$ is the density of the absolutely continuous component 
of $\tau$ with respect to the Lebesgue measure on the unit circle. 
\item[(II)] 
There exists at least one $z$ in the unit disk 
$D=\{z:|z|<1\}$ such that 
 \BEQ{e222}{\sum_{n=0}^\infty|\varphi_n(z)|^2<\infty.} 
\item[(III)] 
There exists at least one $z\in D$ such that 
 $$\liminf_{n\to\infty}|\varphi_n^*(z)|<\infty.$$ 
\item[(IV)] 
The series \eqref{e222} converges uniformly on 
compact subsets of $D$. 
\item[(V)] 
There exists a function $\Pi(z)$, analytic in $D$, 
such that the limit 
\BEQ{e333}{\Pi(z)=\lim_{n\to\infty}\varphi_n^*(z)} 
is uniformly convergent on compact subsets of $D$. 
\end{itemize} 
Moreover, the statements (I\,--V) are equivalent to the 
condition 
$$ 
{\sum_{n=0}^\infty|a_n|^2<\infty.}
$$ 
\end{theorem*}
Note that in (I) the integral is always less than $+\infty$, 
and that there is no restrictions on the singular part of $\tau$.

\section{Krein theorem.\LL{sK}} 

In [{K}] M.~G.~Krein studied the following canonical system of ordinary 
differential equations 
\BEQ{eK}{\begin{aligned} 
\tfrac{d}{dr}& p\, (r,\lambda)\mbox{\hskip-.5em}&=&\,\,
\text{\bfseries\itshape i\/\/} \lambda \, p(r,\lambda) 
- \overline{ a(r)}\,p^*(r,\lambda)\\ 
\tfrac{d}{dr}& p^*(r,\lambda)\mbox{\hskip-.5em}&=&
-{ a(r)}\,p(r,\lambda)
\end{aligned}}
with the initial conditions  
$$p(0,\lambda)=p^*(0,\lambda)=1.$$ 
In our paper we consider only the case when $a(\cdot)$ is 
%a measurable locally integrable function 
continuous on $[0,\infty)$. 
%This system of differential equations is canonical in the sense of Krein. 

There is a 
%unique 
Borel measure $\tau$ on $\mathbb R$, 
which is called the spectral measure, such that 
\BEQ{eee}{\int_{\mathbb R}\frac{1}{1+\lambda^2}d\tau(\lambda)<\infty}
and the map $\mathcal U: L^2_{[0,\infty)}\to L^2_{\tau}$ defined by 
\BEQ{eeI}{\mathcal Uf(\lambda)=\int_0^\infty f(r)p(r,\lambda)dr}
is an isometry. 
%Conversely, if the condition \eqref{eee} is satisfied 
%for a Borel measure $\tau$ on $\mathbb R$, then there exists 
%a unique measurable locally integrable function $a(\cdot)$ on $\mathbb R_+$ 
%such that $\mathcal U$ defined by \eqref{eK} and \eqref{eeI} 
%is an isometry $\mathcal U: L^2_{[0,\infty)}\to L^2_{\tau}$. 

A simple example is the situation when $a(r)\equiv0$ and $\mathcal U$ 
is the usual Fourier transform. In this case 
$\tau$ is the Lebesgue measure normalized by $2\pi$. 
For a more detailed study see [AR,R,D2--5,DK2].

\begin{theorem*} 
The isometry $\mathcal U$ is not onto
 if and only 
if any of the following five equivalent statements hold 
\begin{itemize}
\item[(I)] \BEQ{eK3}
{\int_{\mathbb R}\frac{\log \tau'(\lambda)}{1+\lambda^2} d\lambda 
> -\infty} 
where $\tau'$ is the density of the absolutely continuous component 
of $\tau$ with respect to the Lebesgue measure on $\mathbb R$. 
\item[(II)] 
There exists at least one $\lambda$ in the domain 
$\mathbb C^+=\{\lambda:\text{Im}\lambda>0\}$ such that 
 \BEQ{eK222}{\int_{0}^\infty|p(r,\lambda)|^2dr<\infty.} 
\item[(III)] 
There exists at least one $\lambda\in \mathbb C^+$ such that 
\BEQ{eKinf}{\liminf_{r\to\infty}|p^*(r,\lambda)|<\infty.} 
\item[(IV)] 
The integral \eqref{eK222} converges uniformly on 
compact subsets of $\mathbb C^+$. 
\item[(V)] 
There exists an analytic in $\mathbb C^+$ function $\Pi(\lambda)$ and 
a sequence $r_n\to\infty$ such that the limit 
\BEQ{eK333}{\Pi(\lambda)=\lim_{n\to\infty}p^*(r_n,\lambda)} 
converges uniformly on  compact subsets of $\mathbb C^+$. 
\end{itemize} 
\end{theorem*}

Note that in (I) the integral is always less than $+\infty$, 
and that there is no restrictions on the singular part of $\tau$.

\BREM{remMGK}{This theorem was stated by M.~G.~Krein in 
[{K}] without a proof because of the type of the journal 
it was published in. Parts 
(III,V) of this theorem were not stated in [{K}] correctly. 
Namely, it was written as if (I,II,IV) were equivalent to: 
\begin{itemize}\item[(III$'$)] 
There exists at least one $\lambda\in \mathbb C^+$ such that 
${\sup_{r\geqslant0}|p^*(r,\lambda)|<\infty}$. 
\item[(V$'$)] 
The limit ${\Pi(\lambda)=\lim\limits_{r\to\infty}p^*(r,\lambda)}$ 
converges uniformly on compact subsets of $\mathbb C^+$. 
\end{itemize} 
In Section~\ref{Sce} we 
present two counterexamples. 
We refer to this theorem as the Krein theorem 
because most of the results were stated correctly by M.~G.~Krein in [{K}], 
and the rest is correct in spirit 
despite of a relatively minor mistake.  
The corrected statement 
appeared first in [{T1}].} 
%, along with a proof of the equivalency of (III\,--V). 

 In [{K}]   M.~G.~Krein  noted that if $a\in L^1_{[0,\infty)}$ 
then (I\,--V) hold and $\tau$ is absolutely continuous with positive 
continuous density. Also, he noted that  
if $a\in L^2_{[0,\infty)}$ 
 then (I\,--V) as well as (III',V')  hold. The converse of this fact is not true, 
 unlike the case 
 of orthogonal polynomials in Section~\ref{sPoly}.

In Section~\ref{SBocv} we give a proof that if $a(r)\in L^2[0,\infty)$ 
then (I\,--V) hold, but the 
result is sharp in the sense of Theorem~\ref{thm2} and Remark \ref{remrem}. 
We also prove two more results related to convergence in (I\,--V). 
In Section~\ref{Sce} we prove that, in general, $\Pi(\lambda)$ 
can not be defined uniquely, but only up to a factor 
of absolute value one.

\section{Sakhnovich theorem.\LL{sSakh}} 

In [S1--S5] L.~A.~Sakhnovich introduced and studied matrix analogs 
of the Krein system. He considered a system of canonical 
differential equations 
$${
\tfrac{d}{dr}Y(r,\lambda)=
\text{\bfseries\itshape i\/\/} \lambda J \mathcal H(r)Y(r,\lambda),
\qquad r\geqslant0,}$$
that can be transformed by a change of variables into a system 
\BEQ{eS}{\begin{aligned} 
\tfrac{d}{dr}& P_1^{\phantom*} (r,\lambda) =&& \mbox{\hskip-.5em}
\text{\bfseries\itshape i\/\/} \lambda D P_1^{\phantom*}(r,\lambda) 
+ A_1(r)\,P_1^{\phantom*}(r,\lambda)+ A_2^*(r)\,P_2^{\phantom*}(r,\lambda) \\ 
\tfrac{d}{dr}& P_2^{\phantom*}(r,\lambda) =&& \mbox{\hskip-.5em}
{ A_2^{\phantom*}(r)}\,P_1^{\phantom*}(r,\lambda)
\end{aligned}} 
with the initial conditions 
$$P_1^{\phantom*}(0,\lambda)=P_2^{\phantom*}(0,\lambda)=I_m,$$ 
where $r\in[0,\infty)$, $\lambda\in\mathbb C$, and $I_m$ is the $m\times m$ identity matrix. 
Here $D$, 
$P_1^{\phantom*}(r,\lambda)$, $P_2^{\phantom*}(r,\lambda)$, $A_1(r)$, 
$A_2^{\phantom*}(r)$ are $m\times m$ matrices. 
It is assumed that $A_1(r)=-A_1^*(r)$, and $D$ is a constant 
diagonal matrix with positive diagonal entries. 
Functions $A_1(\cdot)$ and $A_2(\cdot)$ are assumed to be continuous on $[0,\infty)$. 
%locally integrable. 

{There is a 
%unique 
Borel matrix valued measure $\tau$ on $\mathbb R$ such that 
\BEQ{eeeS}{\int_{\mathbb R}\frac{1}{1+\lambda^2}d\tau(\lambda)<\infty}
and the map $\mathcal U: L^2_{[0,\infty)}\to L^2_{\tau}$ defined by 
\BEQ{eeIS}{\mathcal Uf(\lambda)=\int_0^\infty f(r)P_1^{\phantom*}(r,\lambda)dr}
is an isometry.}

\begin{theorem*} 
The following five statements are equivalent 
\begin{itemize}
\item[(I)] \BEQ{eS3}
{\int_{\mathbb R}\frac{\log \det \tau'(\lambda)}{1+\lambda^2} d\lambda 
> -\infty} 
where $\tau'$ is the density of the absolutely continuous component 
of $\tau$ with respect to the Lebesgue measure on $\mathbb R$. 
\item[(II)] 
There exists at least one $\lambda$ in the domain 
$\mathbb C^+=\{\lambda:\text{Im}\lambda>0\}$ such that 
 \BEQ{eS222}{\int_{0}^\infty\|P_1^{\phantom*}(r,\lambda)\|^2dr<\infty,} 
where $\|\cdot\|$ is a matrix norm. 
\item[(III)] 
There exists at least one $\lambda\in \mathbb C^+$ such that 
\BEQ{eSinf}{\liminf_{r\to\infty}\|P_2^{\phantom*}(r,\lambda)\|<\infty.} 
\item[(IV)] 
The integral \eqref{eS222} converges uniformly on 
compact subsets of $\mathbb C^+$. 
\setcounter{footnote}{0}
\item[(V)] 
There exists an analytic in $\mathbb C^+$ matrix valued function $\Pi(\lambda)$ 
 and 
a sequence $r_n\to\infty$ such that the limit 
\BEQ{eS333}{\Pi(\lambda)=\lim_{n\to\infty}P_2^{\phantom*}(r_n,\lambda)} 
converges uniformly on  compact subsets of $\mathbb C^+$. 
\end{itemize} 
\end{theorem*}

\BREM{remLAS}{This important result was proved by L.~A.~Sakhnovich in 
[S2--S4]. 
Unfortunately, 
parts (III,V) of this theorem were not stated in [S2--S4] correctly 
in that it was written as if (I,II,IV) implied the existence of the 
limit \BEQ{ePiP2}{\Pi(\lambda)=\lim_{r\to\infty}P_2^{\phantom*}(r,\lambda).} 
Despite of that, we refer to this theorem as the Sakhnovich 
%Krein 
theorem because most of the results were stated correctly by L.~A.~Sakhnovich, 
and the rest is correct in spirit 
except for a relatively minor mistake.  

The precise location of the gap in Sakhnovich's papers is 
after the proof of the fact 
that $\lim_{n\to\infty}P_1^{\phantom*}(t_n,\lambda)=0$ 
 for a sequence $t_n\to\infty$ (see 
formula (1.35) in \cite{S2} and 
formula (2.10) in \cite{S4}). The  cited formulas do not  
imply  \eqref{ePiP2}. 
What may seem  more surprising is that it does not even imply 
$\lim_{n\to\infty}P_2^{\phantom*}(t_n,\lambda)=\Pi(\lambda)$ but only 
$\lim_{n\to\infty}\|P_2^{\phantom*}(t_n,\lambda)\|=\|\Pi(\lambda)\|$, 
as shown in Theorem~\ref{thm12}.
 
Since the Krein system 
is a particular case of the Sakhnovich system, the counterexamples 
of Section~\ref{Sce} apply to this situation as well. 
Also it is easy to construct ``true'' matrix-valued counterexamples 
along the lines of Section~\ref{Sce}.}

In Section~\ref{SBocv} we show 
that if 
$A_2(r)\in L^2[0,\infty)$, then the finite limit 
$\Pi(\lambda) = \lim_{r\to\infty}P_2^{\phantom*}(r,\lambda)$ exists, and so 
$\Pi(\lambda)$ is unique. 
In Section~\ref{Sce} we prove that, in general, $\Pi(\lambda)$ 
can not be defined uniquely.

%Note that in (I) the integral is always less than $+\infty$, 
%and that there is no restrictions on the singular part of $\tau$. 

Below we give a 
corrected part of the proof of the Sakhnovich theorem. 
Following the lines of [S2--S4], we 
 will show that statements (II\hskip1pt--V) are e\-quiv\-a\-lent. 
An alternative approach can be found in [{T1}]. 

The following lemma is a Lagrange identity,  which is 
an analog of the 
Christoffel-Darboux 
formula for orthogonal 
polynomials (see, for instance, [At]).

\BLEM{lemChristoffel-Darboux}{
\BEQ{eSm}{P_2^*(r,\lambda_0)P_2^{\phantom*}(r,\lambda)-P_1^*(r,\lambda_0)P_1^{\phantom*}(r,\lambda)= 
\text{\bfseries\itshape i\/\/}
(\overline{\lambda_0}-\lambda)\int_{0}^r \mbox{\hskip-.5em} P_1^*(s,\lambda_0)DP_1^{\phantom*}(s,\lambda)ds.}} 
\Pr Note that the relation is clearly true for $r=0$. 
Also, the derivatives with respect to $r$ of both sides of (\ref{eSm}) coincide 
because of (\ref{eS}).  \rP

\Pr[{Proof of a part of the Sakhnovich theorem.}] Statements   
 (II) and (III) are equivalent because of the relation 
\BEQ{eSmm}{P_2^*(r,\lambda)P_2^{\phantom*}(r,\lambda){-}P_1^*(r,\lambda)P_1^{\phantom*}(r,\lambda)= 
2\text{Im}{\lambda}\int_{0}^r \mbox{\hskip-.5em} P_1^*(s,\lambda)DP_1^{\phantom*}(s,\lambda)ds,} 
which is a particular case of (\ref{eSm}). 

Clearly, (IV--V) imply (II\hskip1pt--III) because of  (\ref{eSmm}). 
So we have to show that (II\hskip1pt--III) imply (IV--V). 

Now assume that (II\hskip1pt--III) hold for some $\lambda=\lambda_0\in \mathbb C^+$. 
By (\ref{eS}) and (\ref{eSmm}), the family 
$\{\|P_2^{\phantom*}(r,\lambda)\| : r\geqslant 0, \lambda\in S\}$ 
is uniformly bounded from below for any compact $S\subset \mathbb C^+$. 
By (\ref{eSinf}) and Montel's theorem, there 
exists a sequence $r_n\to\infty$ such that the limit (\ref{eS333}) 
 converges uniformly on compact subsets of~$\mathbb C^+$. 
Thus (V) holds, and so does (IV) because of (\ref{eSmm}). 
\rP

\section{Some convergence results.\LL{SBocv}} 

All the results in this section apply to the Krein system 
if we set 
$m=1$, $D=1$, $A_1(r)=0$, $a(r)=-A_2(r)$, $p(r,\lambda)=P_1^{\phantom*}(r,\lambda)$ and $p^*(r,\lambda)=P_2^{\phantom*}(r,\lambda)$. 

In what follows the matrix norm $\|\cdot\|$ is defined by 
$\|M\|=\sqrt{\mbox{Tr}M^*M}$. 

Note that, even under conditions 
(\ref{thm123}) and (\ref{thm2}) of the following theorem, 
the limit $\lim_{n\to\infty} P_2^{\phantom*}(r_n,\lambda)$ 
may not exist by Remark \ref{remrem}. 

\BTHM{thmSL}{\begin{enumerate}
\item
{\LL{thm123}Suppose that the equivalent conditions 
(I\hskip1pt--V) of the Sakh\-no\-vich theorem hold, and 
$$\lim_{n\to\infty}P_1^{\phantom*}(t_n,\lambda_0)=0$$ for some $t_n\to\infty$ and 
$\lambda_0$ in a nonempty open subset $S$ of $\mathbb C^+$. 
Then the limits
\BEQ{e-limSL}{\begin{aligned}  
&\lim_{n\to\infty} P_2^*(t_n,\xi)P_2^{\phantom*}(t_n,\lambda)=
\Pi^*(\xi)\Pi(\lambda) \\ &\lim_{n\to\infty} \|P_2^{\phantom*}(t_n,\lambda)\|= 
\|\Pi(\lambda)\|
\\ &
\lim_{n\to\infty} P_1^{\phantom*}(t_n,\lambda)=0\end{aligned}  } 
 converge uniformly on compact subsets of $\mathbb C^+\times \mathbb C^+$ and $\mathbb C^+$ 
respectively. Here $\Pi(\lambda)$ is an analytic function on $\mathbb C^+$.} 
\item 
{\LL{thm2}Suppose that the equivalent conditions 
(I\hskip1pt--V) of the Sakhnovich theorem hold, and 
\BEQ{eUI}{\inf\limits_{\varepsilon>0}\Big(\sup\limits_{r\geqslant0}
\int_r^{r+\varepsilon}\|A_2(r)\|dr\Big)=0}
Then the limits
\BEQ{e-limSL-2}{\begin{aligned}  
&\lim_{r\to\infty} P_2^*(r,\xi)P_2^{\phantom*}(r,\lambda)=
\Pi^*(\xi)\Pi(\lambda) \\ &\lim_{r\to\infty} \|P_2^{\phantom*}(r,\lambda)\|= 
\|\Pi(\lambda)\|
\\ &
\lim_{r\to\infty} P_1^{\phantom*}(r,\lambda)=0\end{aligned}  } 
 converge uniformly on compact subsets of $\mathbb C^+\times \mathbb C^+$ and $\mathbb C^+$ 
respectively.} 
\item
{\LL{thmL2}Suppose that 
 $A_2(r)\in L^2[0,\infty)$. 
Then conditions (I\hskip1pt--V) of the Sakhnovich theorem 
hold and, moreover, 
 the limits 
\BEQ{e-lim}{ \begin{aligned}  & 
\lim_{r\to\infty} P_2^{\phantom*}(r,\lambda)=\Pi(\lambda) \\ & 
\lim_{r\to\infty} P_1^{\phantom*}(r,\lambda)=0\end{aligned}  } 
converge uniformly on compact subsets of~$\mathbb C^+$.}
\end{enumerate}} 
 
\BREM{remrem}{This result is sharp in the sense that 
there is a real $C^\infty$ coefficient $A_2(r)$, 
which is in $L^p$ for any $p>2$, 
such that statements (I\hskip1pt--V) of the Sakhnovich 
theorem do not hold. 

Also this result is sharp in a more delicate sense: 
by Theorem~\ref{thm12} there exists a coefficient $A_2(r)$, 
which is again in $L^p$ for any $p>2$, 
such that $ { \lim_ {r\to\infty}} P_1^{\phantom*}(r,\lambda) = 0 $, 
statements (I\hskip1pt--V) of the Sakhnovich theorem hold, but the limit 
$ { \lim_ {r\to\infty}} P_2^{\phantom*}(r,\lambda) $ does not exist.
In fact, we show that 
$\Pi(\lambda)$ can not be defined uniquely, but only up to a constant 
factor, even though the limit 
$ { \lim_ {r\to\infty}} \|P_2^{\phantom*}(r,\lambda)\| = \|\Pi(\lambda)\|$ exists by 
part \ref{thm2} of Theorem~\ref{thmSL}.

Note that in this theorem there is no restriction on the skew-symmetric coefficient 
 $A_1(r)$, 
except for the usual assumption of continuity. 
% local integrability

It was proved in [{S2,S3}] that if 
 $A_2(r)\in L^1[0,\infty)$, 
then conditions (I\hskip1pt--V) of the Sakhnovich theorem 
hold, and the limits (\ref{e-lim}) 
converge uniformly on compact subsets of $\mathbb C^+ \cup \mathbb R$ 
and $\mathbb C^+$ respectively. This fact and 
statement \ref{thmL2} of Theorem~\ref{thmSL} were formulated 
in [{K}] for the Krein system. 
Also, for the Krein system 
statements \ref{thm2} and \ref{thmL2} of Theorem~\ref{thmSL} are 
 related to the results of [{D2}].} 
%closely, which appeared when this work had been completed. 

\Pr[Proof of 1.] 
Identity {\ref{eSm}} implies that if (I\hskip1pt--V) hold and 
\BEQ{eSmpi-y}{\lim_{n\to\infty}P_1^*(r_n,\lambda_0)P_1^{\phantom*}(r_n,\lambda)=0,} 
then $\Pi(\lambda)$ satisfies 
\BEQ{eSmpi}{\Pi(\lambda)= 
\text{\bfseries\itshape i\/\/}
(\overline{\lambda_0}-\lambda)\big(\Pi^*(\lambda_0)\big)^{-1}
\int_{0}^\infty P_1^*(s,\lambda_0)DP_1^{\phantom*}(s,\lambda)ds
.}

Let $\lambda_0\in S$ and $\lambda\in \mathbb C^+$. 
Then using (\ref{eSmm}) at $\lambda_0$ and at $\lambda$ we obtain  
$$\begin{aligned}&P_1^{\phantom*}(t_n,\lambda_0)\mbox{\hskip-1em}&&= o\big( P_2^{\phantom*}(t_n,\lambda_0)\big) _{n\to\infty} ,
\\&
P_1^{\phantom*}(t_n,\lambda)\mbox{\hskip-1em}&&= O\big(P_2^{\phantom*}(t_n,\lambda)\big) 
_{n\to\infty},\end{aligned}$$ 
and therefore 
$$P_1^*(t_n,\lambda_0)P_1^{\phantom*}(t_n,\lambda)= 
o\big( P_2^*(t_n,\lambda_0)P_2^{\phantom*}(t_n,\lambda)\big) 
_{n\to\infty}.$$ 
Hence 
we have (\ref{eSmpi-y}) and  
\BEQ{eSmpi-yy}{\lim_{n\to\infty} 
P_2^*(t_n,\lambda_0)P_2^{\phantom*}(t_n,\lambda)= 
\text{\bfseries\itshape i\/\/}
(\overline{\lambda_0}-\lambda)
\int_{0}^\infty P_1^*(s,\lambda_0)DP_1^{\phantom*}(s,\lambda)ds }  
 by (\ref{eSm}). 

By (\ref{eSmm}) and (\ref{eSmpi-yy}), the family of analytic functions 
$\{P_1^{\phantom*}(t_n,\lambda)\}_{n\geqslant1}$ is locally uniformly bounded 
and so is relatively compact. Thus, any its subsequence has a convergent subsubsequence, and our assumptions imply that its limit has to be zero on $\mathbb C^+$ because 
it is an analytic function which is zero on an nonempty open set $S$. Hence  
$\lim_{n\to\infty} P_1^{\phantom*}(t_n,\lambda)=0$  uniformly on compact subsets of~$\mathbb C^+$. 

Therefore the sequence $\{P_2^{\phantom*}(t_n,\lambda)\}_{n\geqslant1}$ is 
%locally uniformly 
bounded by (\ref{eSmm}), (\ref{eSmpi-y}),  and so has a convergent subsequence.  
%, say $\{P_2^{\phantom*}(t_{n_k},\lambda)\}_{n\geqslant1}$
Then  we  define $\Pi(\lambda)$ as the limit of this  
subsequence. The right hand side of (\ref{eSmpi-yy}) does not 
depend on the choice of the subsequence, and so (\ref{eSmpi-yy}) 
extends to $\lambda_0,\lambda\in \mathbb C^+$ by analyticity. 
This implies the first and second limits in (\ref{e-limSL}). 
\rP

\Pr[Proof of 2.] From (\ref{eS}) we have that 
\BEQ{eSlog-0}{
\begin{aligned}&
\tfrac{d}{dr} \|P_1^{\phantom*} (r,\lambda)\|^2=
\mbox{Tr} \tfrac{d}{dr} P_1^{\smash*} P_1^{\phantom*}   =\\&
\mbox{Tr} 
\big({
{-}2\text{Im}{\lambda} P_1^*        D P_1^{\phantom*}        + 
P_1^* A_2^{\phantom*}  P_2^{\phantom*}       +
P_2^* A_2^{\phantom*}  P_1^{\phantom*}
}\big)
\geqslant \\&
-2\big(\text{Im}{\lambda}\|D\| \|P_1^{\phantom*} (r,\lambda)\|^2 +
\|A_2^{\phantom*} (r)\| \|P_1^{\phantom*} (r,\lambda)\|\|P_2^{\phantom*} (r,\lambda)\|\big)
\end{aligned}
} 
and also 
\BEQ{eSlog}{
\begin{aligned}&
\tfrac{d}{dr} 
\log \big(\|P_1^{\phantom*} (r,\lambda)\|^2+\|P_2^{\phantom*} (r,\lambda)\|^2\big)=
\mbox{Tr} \frac
{\frac{d}{dr} 
\big( P_1^*        P_1^{\phantom*}        +P_2^*       P_2^{\phantom*}       \big)}
{ \|P_1^{\phantom*}       \|^2+\|P_2^{\phantom*}       \|^2}=\\&
2\mbox{Tr} \frac
{
{-}\text{Im}{\lambda} P_1^*        D P_1^{\phantom*}        + 
P_1^* A_2^{\phantom*}  P_2^{\phantom*}       +
P_2^* A_2^{\phantom*}  P_1^{\phantom*}
}
{ \|P_1^{\phantom*}       \|^2+\|P_2^{\phantom*}       \|^2}
\leqslant 4\|A_2^{\phantom*} (r)\| 
\end{aligned}
} 
since $\mbox{Im}\lambda>0$. 

Let us assume that $\limsup_{r\to\infty}\|P_1^{\phantom*}(r,\lambda)\|>0$ for some 
$\lambda\in \mathbb C^+$. Then 
there is a sequence $t_n\to\infty$ such that 
$$\lim_{n\to\infty}\|P_1^{\phantom*}(t_n,\lambda)\|=\delta>0.$$ 
Relation (\ref{eSmm}) implies that 
$$\lim_{n\to\infty}\|P_2^{\phantom*}(t_n,\lambda)\|=\gamma>\delta.$$ 
Then (\ref{eSm}) and (\ref{eSlog}) implies that for any $\delta_0,\varepsilon_0>0$  there exist 
$C>0$ such that
$$\|P_1^{\phantom*} (r,\lambda)\|^2+\|P_2^{\phantom*} (r,\lambda)\|^2\leqslant 
(\delta^2+\gamma^2)\exp\Big(
\int_{t_n}^{t_n+\varepsilon_0}
\mbox{\hskip-1em}
4\|A_2^{\phantom*}(r)\|dr\Big)+\delta_0
<C$$ 
 for all large enough $n$ and any $r\in[t_n,t_n+\varepsilon_0]$. 
Therefore we can conclude from  (\ref{eSlog-0}) that there 
 are $\delta_1>0$ and $\varepsilon_1>0$ such that 
$$\|P_1^{\phantom*} (r,\lambda)\|>\delta_1$$ 
 for all large enough $n$ and any $r\in[t_n,t_n+\varepsilon_1]$. 
This is a contradiction with (\ref{eS222}), 
and so $\lim_{r\to\infty} P_1^{\phantom*}(r,\lambda) = 0$ for any 
$\lambda\in \mathbb C^+$. 

Then the proof of (\ref{e-limSL-2}) 
follows from statement (1) of this theorem. 
\rP

\Pr[Proof of 3.]
Our first aim is to show that 
the integral (\ref{eS222}) converges for any $\lambda\in \mathbb C^+$. 
Let us assume that $\int_{0}^\infty\|P_1^{\phantom*}(r,\lambda)\|^2dr=\infty$ for some 
$\lambda\in \mathbb C^+$. Then one can see that 
$$
\|P_2^{\phantom*}(r,\lambda)\|^2 \leqslant 
{\Big(\int_{0}^r \|A_2^{\phantom*}(s)\|\|P_1^{\phantom*}(s,\lambda)\|ds\,\Big)^2= 
o\Big(\int_{0}^r \|P_1^{\phantom*}(s,\lambda)\|^2ds\Big)_{r\to\infty},} 
$$
which contradicts to (\ref{eSmm}). Thus 
the integral (\ref{eS222}) converges for any $\lambda\in \mathbb C^+$ and so 
$${
\Pi(\lambda)=\lim_{r\to\infty}P_2^{\phantom*}(r,\lambda) 
=I_m+\int_{0}^\infty A_2^{\phantom*}(r) P_1^{\phantom*}(r,\lambda)dr
}$$
 holds, 
since $A_2^{\phantom*}(r)\in L^2[0,\infty)$. 
The rest of the proof follows from (\ref{eSmm}), (\ref{eSm}). 
\rP

\section{Two results on nonconvergence.\LL{Sce}}

\BTHM{thm111} 
{There exists a real-valued continuous function $a(r)$ such that 
the spectral measure $\tau$ is absolutely continuous 
with positive continuous density, 
statements (I\,--V) of the Krein theorem hold, but 
\BEQ{eKthm111}{\liminf_{r\to\infty}|p^*(r,\lambda)| 
<\limsup_{r\to\infty}|p^*(r,\lambda)|} 
for any $\lambda\in \mathbb C^+$. 
In addition, 
the $\limsup$ in (\ref{eKthm111}) can be either finite 
or identically $+\infty$ on $\mathbb C^+$.} 

\BREM{remthm111} 
{In this theorem, by construction, $a(r)$ can be chosen to be a $\,C^\infty$ function.} 

Before giving a detailed proof of Theorem \ref{thm111}, we  
describe 
a simple construction of a function $a(r)$ such that 
\eqref{eKthm111} holds for a fixed 
$\lambda\in \mathbb C^+$. 
\Pr[A sketch of the proof of Theorem \ref{thm111}] 
We choose positive constants $\varepsilon_n$ and 
$r_n$ such that $\varepsilon_n\to0$ and 
$r_n-r_{n-1}\to\infty$ as $n\to\infty$, and then define 
\begin{equation*}
a(r)=\left\{\begin{aligned} 
-\frac{1}{\varepsilon_n},&\ \ r\in[r_n,r_n{+}\varepsilon_n)\\ 
\frac{1}{\varepsilon_n},&\ \ r\in[r_n{+}\varepsilon_n,r_n{+}2\varepsilon_n)\\ 
0,&\ \ r\in[r_n{+}2\varepsilon_n,r_{n+1})
\end{aligned}\right.
\end{equation*}
assuming the intervals involved do not intersect each other and 
$r_0=0$. 
Note that  $p^*(r,\lambda)$ is constant and 
$|p(r,\lambda)|$ decreases 
exponentially when $r\in[r_n{+}2\varepsilon_n,r_{n+1})$. 
So we can assume $|p(r_n,\lambda)|$ are arbitrarily small 
%(tend to zero arbitrarily fast) 
if $r_n-r_{n-1}$ are large enough. 
Then it is easy to see that, if $\varepsilon_n$ are small enough, 
 $p^*(r_n{+}\varepsilon_n,\lambda)$ are  arbitrarily close to 
 $\cosh{(1)}\,p^*(r_n,\lambda)$  and 
 $p^*(r_n{+}2\varepsilon_n,\lambda)$ are  arbitrarily close to 
 $p^*(r_n,\lambda)$.
To justify it formally, see \eqref{eQ} and consider 
the change of variable $s=r/\varepsilon_n$. 
Thus,  if 
 $r_n-r_{n-1}$ are large enough and 
$\varepsilon_n$ are small enough, then 
$\liminf_{r\to\infty}|p^*(r,\lambda)|$
is arbitrarily close to $1$ and 
$\limsup_{r\to\infty}|p^*(r,\lambda)|$
is arbitrarily close to $\cosh{(1)}$. 
 \rP

Before the proof of Theorem \ref{thm111}, we need the following lemma.

\BLEM{lem1}{Let $b(r)$ be any real continuous 
function such that 
$$\int_0^1 b(r)dr=0.$$ 
For $0<\varepsilon<1$ let  
 $p_\varepsilon (r,\lambda)$ and $p^*_\varepsilon (r,\lambda)$ be the solutions 
of (\ref{eK}) with 
$$a(r)=a_\varepsilon(r)=-
\tfrac{\log|\log\varepsilon|}{\varepsilon}b(\tfrac{r}{\varepsilon})$$ 
and initial conditions $p_\varepsilon (0,\lambda)=c$, 
$p^*_\varepsilon (0,\lambda)=c^*$. 

Then 
\BEQ{eKcl2}{\begin{aligned}
&p_\varepsilon(\varepsilon,\lambda)
=c\mbox{\hskip-.9em}&+o(\sqrt{\varepsilon})_{\varepsilon\to0}
 \\
&p^*_\varepsilon(\varepsilon,\lambda)
=c^*\mbox{\hskip-.9em}&+o(\sqrt{\varepsilon})_{\varepsilon\to0}\end{aligned}} 
where the limits are uniform for $\lambda$,
$c$,
$c^*$ in any compact subset of $\mathbb C$. In addition, if 
$c\neq -c^*$ and 
\BEQ{eKpos}{\int_0^{\frac12} b(r)dr>0,} 
then 
\BEQ{eKcl}{
\lim_{\varepsilon\to0}|p_\varepsilon(\tfrac\varepsilon2,\lambda)|=
\lim_{\varepsilon\to0}|p^*_\varepsilon(\tfrac\varepsilon2,\lambda)|=\infty
.}} 

\Pr
First, we consider differential equations 
\BEQ{eKqq}{\begin{aligned} 
\tfrac{d}{dr}&\, q_\varepsilon (r)&\,=\,& 
- { a_\varepsilon(r)}\,q^*_\varepsilon(r)\\ 
\tfrac{d}{dr}&\, q^*_\varepsilon(r)&\,=\,& 
-{ a_\varepsilon(r)}\,q_\varepsilon(r)
\end{aligned}}
with initial conditions 
$q_\varepsilon (0)=c$, $q^*_\varepsilon (0)=c^*$. 
Then we have 
\BEQ{eKqqq}{\begin{aligned} 
q_\varepsilon(r)-q^*_\varepsilon(r)=\,&
(c-c^*)\exp\Big\{\int_0^r a_\varepsilon(r)dr\Big\},
\\ 
q_\varepsilon(r)+q^*_\varepsilon(r)=\,&
(c+c^*)\exp\Big\{-\int_0^r a_\varepsilon(r)dr\Big\}
 .
\end{aligned}} 
Hence $
q_\varepsilon(\varepsilon)=c$ 
and 
$q^*_\varepsilon(\varepsilon)=c^*$. 
Thus our aim is to show that for $0\leqslant r\leqslant\varepsilon$ we have 
$$
|p_\varepsilon(r,\lambda)-q_\varepsilon(r)|=
o(\sqrt{\varepsilon})_{\varepsilon\to0}
 \mbox{\qquad and\qquad}
|p^*_\varepsilon(r,\lambda)-q^*_\varepsilon(r)|=
o(\sqrt{\varepsilon})_{\varepsilon\to0}
.$$ 

To show this, we use Gronwall's lemma: \emph{if $\alpha(r)$ is a nonnegative 
integrable function such that 
\BEQ{eGron1}{\alpha(r)\leqslant c_1\int_0^r\alpha(s)ds+c_2} 
for some constants $c_1,c_2\geqslant0$,  then $$\alpha(r)\leqslant c_2e^{c_1r}.$$} 

First, we  use Gronwall's lemma with 
$$c_1=M_\varepsilon=|\lambda|+ 
\tfrac{\log|\log\varepsilon|}{\varepsilon} 
\max\limits_{0\leqslant s\leqslant1}|b(s)|$$
and
$c_2=|c|+|c^*|$ 
 to estimate 
$\alpha(r)=|p_\varepsilon(r,\lambda)|+
|p^*_\varepsilon(r,\lambda)|$. Thus, by (\ref{eK}) and the definition of 
$p_\varepsilon(r,\lambda)$ and $p^*_\varepsilon(r,\lambda)$ we have 
\BEQ{eKGW}{|p_\varepsilon(r,\lambda)|+
|p^*_\varepsilon(r,\lambda)|\leqslant(|c|+|c^*|)e^{M_\varepsilon r}.} 
Then we use Gronwall's lemma once more to estimate 
$$\alpha(r)=|p_\varepsilon(r,\lambda)-q_\varepsilon(r)|+ 
|p^*_\varepsilon(r,\lambda)-q^*_\varepsilon(r)|.$$ 
Using the previous estimate, (\ref{eK}) and  (\ref{eKqqq})   we obtain 
(\ref{eGron1}) with $c_1=M_\varepsilon$ and 
$$c_2=\varepsilon|\lambda|(|c|+|c^*|)e^{M_\varepsilon\varepsilon}\geqslant r |\lambda\,p_\varepsilon(s,\lambda)| $$ 
for any $0\leqslant r\leqslant\varepsilon$. 
Then by estimate (\ref{eKGW}) we have 
$$ 
|p_\varepsilon(r,\lambda)-q_\varepsilon(r)|+ 
|p^*_\varepsilon(r,\lambda)-q^*_\varepsilon(r)|\leqslant 
\varepsilon |\lambda| (|c|+|c^*|)e^{2M_\varepsilon\varepsilon } 
=o(\sqrt\varepsilon)_{\varepsilon\to0}$$ 
for any $0\leqslant r\leqslant\varepsilon$. 

Moreover, by (\ref{eKpos}) and (\ref{eKqqq}) 
%if $c\neq -c^*$ then 
$$
q_\varepsilon(\tfrac\varepsilon2)=
q^*_\varepsilon(\tfrac\varepsilon2)+o(1)_{\varepsilon\to0}=
\frac12(c+c^*)\exp\Big\{{\log|\log\varepsilon|}\cdot\int_0^{\frac12} b(r)dr\Big\}
+o(1)_{\varepsilon\to0},$$
which completes the proof. 
\rP

\Pr[Proof of Theorem~\ref{thm111}] 
In this proof 
$n\to\infty$ means that the limit is taken over positive integers, and 
$r\to\infty$ means that the limit is taken over positive reals. 

We fix a function $b(r)$ which 
satisfies all the conditions of Lemma \ref{lem1}. 
Also we assume that  $b(r)=0$ if $r\notin[0,1]$. 
Let $a(r)$ be defined by 
$$ 
a(r)=-\sum_{n=1}^\infty (2^n\log n) b(2^nr-n2^n)= 
\sum_{n=1}^\infty a_{\varepsilon_n}(r-n), 
$$ 
where $a_\varepsilon(\cdot)$ is defined as in Lemma \ref{lem1}, and 
${\varepsilon_n}=2^{-n}$. This sum is a continuous  
function since for any $r$ the sum contains at most one 
nonzero term. 
Then by Lemma \ref{lem1} we have 
$$
|p^*(n,\lambda)-p^*(n+2^{-n},\lambda)|=
o(2^{-n/2})_{n\to\infty}.
$$
Note that $p^*(r,\lambda)$ does not change when $r$ is in an interval 
$[n+2^{-n},n+1]$ since $a(r)=0$ on such intervals. Therefore  by (\ref{eK}) we have   
\BEQ{ePS}{
|p^*(n,\lambda)-p^*(n+1,\lambda)|=
o(2^{-n/2})_{n\to\infty}.}
Hence the limit $\lim_{n\to\infty}p^*(n,\lambda)$ exists and is finite 
for any $\lambda\in\mathbb C$. 
Note that $\lim_{n\to\infty}p^*(n,\lambda)\neq0$ for 
$\text{Im}\lambda\geqslant0$ 
since, by (\ref{eK}), 
\BEQ{eKnz}{ 
\frac{d}{dr}\left(|p^*(r,\lambda)|^2-|p(r,\lambda)|^2\right)=
2\text{Im}\lambda|p(r,\lambda)|^2\geqslant0.} 
By the same argument, for any $r>0$ and $\text{Im}\lambda>0$ 
we have 
%$\lim_{n\to\infty}p(n,\lambda)=0$ 
%since 
%$\frac{d}{dr}p(r,\lambda)=\text{\bfseries\itshape i\/\/} \lambda p(r,\lambda)$ 
%on any interval $[n+2^{-n},n+1]$. Hence
 $p(r,\lambda)\neq -p^*(r,\lambda)$. Then 
Lemma \ref{lem1} implies that 
$$\lim_{n\to\infty}|p(n+2^{-n-1},\lambda)|=
\lim_{n\to\infty}|p^*(n+2^{-n-1},\lambda)|=
\infty. $$
Note that if in Lemma \ref{lem1} 
we define 
$a_\varepsilon(r)=-
\tfrac{M}{\varepsilon}b(\tfrac{r}{\varepsilon})$, then 
 $$\liminf_{r\to\infty}|p^*(r,\lambda)|
<\limsup_{r\to\infty}|p^*(r,\lambda)|<\infty $$ 
for any large enough $M$.

In order to complete the proof we need to show that 
the spectral measure $\tau$ is absolutely continuous 
with positive continuous density. 
The estimates (\ref{ePS}) and Lemma \ref{lem1} shows that the limit 
$\Pi(\lambda)=\lim_{n\to\infty}p^*(n,\lambda)$ 
converges uniformly on compact sets of $\lambda\in\mathbb C$. 
As a byproduct we have proved that $\Pi(\lambda)$ is continuous 
for $\lambda\in\mathbb C$ 
and has no zeros in the closed half-plane 
$\mbox{Im}\lambda\geqslant0$. 
In particular, this is so for real $\lambda$. 

For the rest of the proof 
we assume $\lambda\in\mathbb R$. 
Let $\tau_r$ be the measure absolutely continuous with respect to 
the Lebesgue measure with the density 
$$ 
\frac{d\tau_r(\lambda)}{d\lambda}=\frac1{2\pi|p^*(r,\lambda)|^2}.
$$ 
Then $\tau_r$ converges weakly to $\tau$ as $r\to\infty$
(see, for instance, [{T1}]). By the previous paragraph, 
$$\frac{d\tau(\lambda)}{d\lambda}=
\lim_{n\to\infty}\frac1{2\pi|p^*(n,\lambda)|^2}=
\frac1{2\pi|\Pi(\lambda)|^2}$$ is a positive continuous function 
on $\mathbb R$, which completes the proof. 
\rP

\BTHM{thm12}{There exists a continuous  
function $a(r)$ such that (I\,--V) of the Krein theorem hold, but 
the function 
$\Pi(\lambda)$, which is analytic in $\mathbb C^+=\{\lambda:\text{Im}\lambda>0\}$, 
 is not unique in the following sense: 
for any complex\/ $\theta$ of absolute value one there is a sequence 
$t_n\to\infty$ such that 
\BEQ{eKthm1}{\lim_{n\to\infty}p^*(t_n,\lambda)=\theta\Pi(\lambda).}
In addition, we can have the following conditions satisfied:  
$a(r)\in L^p[0,\infty)$ for any $p>2$,  
$\lim_{r\to\infty}a(r)=0$, 
and for any $\lambda\in \mathbb C^+$
\BEQ{eGG}{\begin{aligned} &
\lim_{r\to\infty}p(r,\lambda)=0 \\ & 
\lim_{r\to\infty}|p^*(r,\lambda)|=|\Pi(\lambda)|.\end{aligned} 
}} 
 
\BREM{remthm1112} 
{In this theorem, by construction, $a(r)$ can be chosen to be a $\,C^\infty$ function.}

\Pr 
We will construct a function $a(r)$ which is piecewise constant, 
and then can be approximated by 
continuous functions that still have the desired properties. 

First, note that the system of differential equations 
\BEQ{eKqq2}{\begin{aligned} 
\tfrac{d}{dr}&\, q (r) &\mbox{\hskip-.8em}=\,& 
- \overline{a(r)}\,q^*(r)\\ 
\tfrac{d}{dr}&\, q^*(r)&\mbox{\hskip-.8em}=\,& 
-{a(r)}\,q(r)
\end{aligned}} 
with constant coefficient $a(r)=-C$ 
has a matrix solution 
\BEQ{eQ}{Q(r)=\begin{pmatrix}
  &\cosh &\mbox{\hskip-.7em}|Cr| & {{\overline D}} \mbox{\hskip-0.75em}&\sinh &\mbox{\hskip-.7em}|Cr|
\\
D \mbox{\hskip-.75em}&\sinh &\mbox{\hskip-.7em}|Cr| &                  &\cosh &\mbox{\hskip-.7em}|Cr|
\end{pmatrix}}
where $D=\frac{C}{|C|}$. 

Now let $b$ be positive real and 
\BEQ{eKqqaa}{a_{b,\xi,\varepsilon}(r)=\left\{
\begin{aligned} 
-b, & \mbox{ \ for} \ 0\leqslant r\leqslant\varepsilon , \cr 
\overline\xi b, & \mbox{ \ for} \ 
\varepsilon\leqslant r\leqslant2\varepsilon , \cr 
0, & \mbox{ \ for} \ r\geqslant2\varepsilon, \cr 
\end{aligned}\right.} 
 where the constant $\xi\in\mathbb C$ is such that $|\xi|=1$. 
Let $q(r)=q_{b,\xi,\varepsilon}(r)$ and 
$q^*(r)=q_{b,\xi,\varepsilon}^*(r)$ be 
the solutions of the system of equations (\ref{eKqq2}) 
with $a(r)=a_{b,\xi,\varepsilon}(r)$, and initial conditions 
$q (0)=0$, $q^* (0)=1$. 
Then 
\BEQ{eKqq23}{\begin{aligned} 
q _{b,\xi,\varepsilon} (\varepsilon)& =\sinh b\varepsilon, 
& \qquad & \ & \ & \ & \ & \ &q _{b,\xi,\varepsilon} (2\varepsilon)
&= \frac12(1-\xi) 
\sinh 2b\varepsilon, \\ 
q_{b,\xi,\varepsilon}^* (\varepsilon)& =\cosh b\varepsilon, 
& \qquad & \ & \ & \ & \ & \ &q_{b,\xi,\varepsilon}^* (2\varepsilon)
&= 1+(1-\overline\xi)\sinh^2 b\varepsilon. 
\end{aligned}} 
Let $p_{b,\xi,\varepsilon}(r,\lambda)$ and 
$p^*_{b,\xi,\varepsilon}(r,\lambda)$ be the solutions of 
 the system of equations (\ref{eK}) with 
$a(r)=a_{b,\xi,\varepsilon}(r)$, and initial conditions 
$p_{b,\xi,\varepsilon}(0,\lambda)=0$ and 
$p^*_{b,\xi,\varepsilon}(0,\lambda)=1$.

To estimate these solutions we use 
the following form of Gronwall's lemma: 
\emph{if $\alpha(r)$ is a nonnegative 
integrable function such that
\BEQ{eGron2}{\alpha(r)\leqslant c\int_0^r\alpha(s)ds 
+\beta(r)}  for some 
$c$ and $\beta(r)\geqslant0$, then 
\BEQ{eGron22}{\alpha(r)\leqslant 
c\int_0^re^{c(r-s)}\beta(s)ds+\beta(r).}}

In the following estimates we assume that $\lambda\in\mathbb C$ is fixed. 
We write ``$const$'' for a constant, different in different inequalities, 
which depends on $\lambda$, but is independent of 
 $\varepsilon$, $r$ and $b$ provided $0<\varepsilon,r,b<1$. 

First, we use Gronwall's lemma 
with $\alpha(r)=|p_{b,\xi,\varepsilon}(r,\lambda)|+
|p^*_{b,\xi,\varepsilon}(r,\lambda)|$. Then (\ref{eK}) implies (\ref{eGron2}) 
with $\beta(r)=1$ and $c=|\lambda|+ b$
and  so (\ref{eGron22}) implies 
$$
|p_{b,\xi,\varepsilon}(r,\lambda)|+
|p^*_{b,\xi,\varepsilon}(r,\lambda)|\leqslant 
e^{(|\lambda|+ b) r}
<const.$$
Second, we apply this form of Gronwall's lemma 
with $\alpha(r)=|p_{b,\xi,\varepsilon}(r,\lambda)|$. 
Then (\ref{eK}) and the previous estimate imply (\ref{eGron2}) 
with  $c=|\lambda|$ and 
$$\beta(r)=const\cdot br > \int_0^r|b\,p^*_{b,\xi,\varepsilon}(s,\lambda)|ds.$$ 
Therefore (\ref{eGron22}) implies 
$$
|p_{b,\xi,\varepsilon}(r,\lambda)|
<const\cdot br. 
$$

Using the same form of Gronwall's lemma the third time  
with $c=|\lambda|+ b$, 
$$\beta(r)=const\cdot br^2 > \int_0^r|\lambda\,p_{b,\xi,\varepsilon}(s,\lambda)|ds$$ 
and  
$$\alpha(r)=|p_{b,\xi,\varepsilon}(r,\lambda)-q_{b,\xi,\varepsilon}(r)|+
|p^*_{b,\xi,\varepsilon}(r,\lambda)-q_{b,\xi,\varepsilon}^*(r)|,$$ 
we obtain 
\BEQ{eEST1}{
|p_{b,\xi,\varepsilon}(r,\lambda)-q_{b,\xi,\varepsilon}(r)|+
|p^*_{b,\xi,\varepsilon}(r,\lambda)-q_{b,\xi,\varepsilon}^*(r)|<
const\cdot br^2} 
by (\ref{eK}), (\ref{eKqq2}) and the previous estimates. This implies 
\BEQ{eEST2}{
|p_{b,\xi,\varepsilon}^*(r,\lambda)-q_{b,\xi,\varepsilon}^*(r)|<
const\cdot b^2r^3
} 
by (\ref{eK}) and (\ref{eKqq2}). 

We define 
$$
\varepsilon_n=\frac{1}{\log^2 n}, \qquad b_n=\frac{\log^2 n}{\sqrt n} 
$$
for $n\geqslant3$. Also we define $\xi_n$ as a unique complex number 
such that 
$$|\xi_n|=1, \qquad |1-\xi_n|=\frac{1}{\log n} 
\qquad \mbox{ and } \qquad\mbox{Im}\xi_n>0.$$ 
Note that 
\BEQ{eXi}{\xi_n=1+
\dfrac{\text{\bfseries\itshape i\/\/}}{\log n}
+O\Big(\frac{1}{\log^2 n}\Big)_{n\to\infty}. }

Let $a(r)$ be defined by 
$$ 
a(r)= 
\sum_{n=3}^\infty a_{b_n,\xi_n,\varepsilon_n}(r-r_n), 
$$ 
where $a_{b,\xi,\varepsilon}(\cdot)$ is defined by (\ref{eKqqaa}), and 
$r_n$ are as follows. We fix any $\lambda_0\in \mathbb C^+$. 
Then we choose $r_2=0$ and each $r_n-r_{n-1}$ 
to be large enough so that 
\BEQ{eEST123}{
\frac{p^*(r_n+2\varepsilon_n,\lambda_0)}{p^*(r_n,\lambda_0)} = 1 + 
\frac{\text{\bfseries\itshape i\/\/}}{ n\log n} + 
O\left(\frac{1}{ n\log^2 n}\right)_{n\to\infty}.} 
This is possible 
 since $|p(r_{n},\lambda_0)|\to0$ exponentially 
as $r_{n-1}$ is fixed and $(r_n-r_{n-1})\to\infty$. 
Therefore we can use (\ref{eEST2}), (\ref{eXi}), and the fact that 
$$
{q_{b,\xi,\varepsilon}^*(2\varepsilon,\lambda)} 
= 1+(1-\overline\xi) \Big(b^2 \varepsilon^2 
+ O(b^4 \varepsilon^4)_{b\varepsilon\to0}\Big)
$$ 
by (\ref{eKqq23}). 

We have that $p^*(r,\lambda)$ is constant for 
$r\in[r_n+2\varepsilon_n,r_{n+1}]$, in particular, $$p^*(r_n+2\varepsilon_n,\lambda)=p^*(r_{n+1},\lambda) . $$
Hence (\ref{eEST123}) imply that 
$$
\left|\frac{p^*(r_{n+1},\lambda_0)}{p^*(r_n,\lambda_0)}\right|-1
=
O\left( \frac{1}{ n \log^2 n}\right)_{n\to\infty}
$$ 
and so the limit 
$\lim_{n\to\infty}|p^*(r_n,\lambda_0)|=|\Pi(\lambda_0)|$ 
converges, since 
$$\sum_{n=3}^\infty\frac1{n\log^2n}<\infty.$$ 
Thus statements (I\,--V) of the Krein theorem hold by 
(\ref{eKK}) and (\ref{eKinf}).

If each $r_n-r_{n-1}$ 
is large enough, then the sum that defines $a(r)$ is a sum of the 
functions with disjoint support. Therefore  
 $$\|a(r)\|_{L^p}^p=2\sum_{n=3}^\infty n^{-p/2}\log^{2p-2} n,$$ 
and so $a(r)\in L^p[0,\infty)$ if and only if $p>2$. 
In particular, this means that  part (\ref{thm2}) of Theorem~\ref{thmSL} implies (\ref{eGG}). 

To complete the proof note that 
the limit $\lim_{n\to\infty}p^*(r_n,\lambda_0)$ does not exists 
because \BEQ{eEST12345}{
\frac{p^*(r_n+2\varepsilon_n,\lambda_0)}{p^*(r_n,\lambda_0)} = \exp\Big\{
\frac{\text{\bfseries\itshape i\/\/}}{ n\log n} + 
O\left(\tfrac{1}{ n\log^2 n}\right)_{n\to\infty}\Big\}} 
by (\ref{eEST123}), and the series 
$\sum_{n=3}^\infty\frac1{n\log n}$ diverges, while $\sum_{n=3}^\infty\frac1{n\log^2 n}<\infty$. 
At the same time 
$\lim_{n\to\infty}\frac1{n\log n}=0$
and so for any complex $\theta$ of absolute value one there is a sequence 
$t_{\theta,n}\to\infty$, which is a subsequence of $r_n$, such that 
$$\lim_{n\to\infty}p^*(t_{\theta,n},\lambda_0)=\theta|\Pi(\lambda_0)|. $$

Note that $|\Pi(\lambda)|$ is well defined for any $\lambda\in \mathbb C^+$ 
since $\lim_{r\to\infty}|p^*(r,\lambda)|=|\Pi(\lambda)|$ 
converges by (\ref{e-limSL-2}). 
Also using (\ref{e-limSL-2}) we can define a function 
$\Pi(\lambda)$, which is analytic in $\mathbb C^+$, by 
$$\Pi(\lambda)=|\Pi(\lambda_0)|^{-1} 
\lim_{n\to\infty}p^*(t_{1,n},\lambda) 
\overline{p^*(t_{1,n},\lambda_0)}= \lim_{n\to\infty}p^*(t_{1,n},\lambda) 
.$$ 
Then 
$\lim_{n\to\infty}p^*(t_{\theta,n},\lambda)=\theta\Pi(\lambda)$ 
for any $\lambda\in \mathbb C^+$ because of (\ref{e-limSL-2}). 
\rP 

\BB{proposition}{prop1}{If $r_n-r_{n-1}$ are large enough 
in the proof of Theorem \ref{thm12}, then 
for all $\lambda\in \mathbb C^+$ we have 
(\ref{eEST123}) as well as estimates 
\BEQ{eEST144}{ 
\left|\frac{p(r,\lambda)}{p^*(r_n,\lambda)}\right| 
< \frac{const}{\sqrt n\log n}} 
for $r_n+2\varepsilon_n\leqslant r \leqslant r_{n+1}$, and 
\BEQ{eEST12}{ 
\left|\frac{p^*(r,\lambda)}{p^*(r_n,\lambda)}-1\right|
<\frac{const}{ n }, 
\qquad 
\qquad 
\left|\frac{p(r,\lambda)}{p^*(r_n,\lambda)}\right|
< \frac{const}{\sqrt n}} 
for $r_n\leqslant r \leqslant r_n+2\varepsilon_n$. 
This gives, in particular, 
a constructive proof of (\ref{eK222}) and (\ref{eGG}).} 

\Pr 
We can  demonstrate (\ref{eEST144}) and (\ref{eEST12}) 
for $\lambda=\lambda_0$ using estimates 

$$
\left|{q_{b,\xi,\varepsilon}^*(r,2\varepsilon)}\right|< 
const\cdot b\varepsilon |1-\xi|
$$
and, for $0\leqslant r \leqslant 2\varepsilon$, 
$$
\left|{q_{b,\xi,\varepsilon}(r,\lambda)}\right| 
< const\cdot b \varepsilon, 
\qquad\qquad
\left|{q_{b,\xi,\varepsilon}^*(r,\lambda)}-1\right| 
< const\cdot b^2 \varepsilon^2 
$$ 
which follows from \ref{eQ} and \ref{eKqq23}. 

We also can obtain (\ref{eEST12}) and (\ref{eEST123}) 
for all $\lambda\in \mathbb C^+$
 if the sequence $r_n$ is 
chosen as follows. It is easy to see that estimates like 
(\ref{eEST1}) and (\ref{eEST2}) can be established uniformly 
in $\lambda$ in a compact subsets of $\mathbb C$. 
Also 
 $|p(r_{n},\lambda)|\to0$ uniformly 
in $\lambda$ in a compact subsets of $\mathbb C^+$ 
as $r_{n-1}$ is fixed and $(r_n-r_{n-1})\to\infty$. Thus 
for any compact subset $H$ of $\mathbb C^+$ there is a sequence 
$r^H_n$ such that (\ref{eEST123}), (\ref{eEST144}) and (\ref{eEST12}) 
hold for 
$r_n=r^H_n$, and also for $r_n$ that is any subsequence of $r^H_n$. 
We can represent $\mathbb C^+$ as an increasing union of 
compact subsets $H_k$. Without loss of generality we can assume that 
$r^{H_{k+1}}_n$ is a subsequence of $r^{H_k}_n$ for each $k$. 
Then we define $r_n$ by the ``diagonal process'' $r_n=r^{H_{n}}_n$. 
\rP

\BB{conjecture}{con1}{We conjecture that if $a(r)$ 
is a \emph{real-valued} function, and conditions (I\hskip1pt--V) of the Krein 
theorem hold, 
then $\Pi(\lambda)$ is unique in the following sense: 
if $t_n\to\infty$ 
and $\lim_{n\to\infty}p(t_n,\lambda)=0$, then the limit 
$\lim_{n\to\infty}p^*(t_n,\lambda)=\Pi(\lambda)$ 
converges uniformly on compact subsets of~$\mathbb C^+$.
If true, this conjecture implies that the original form of 
Krein's theorem holds if $a(r)$ 
is real and ``locally uniformly integrable'' in the sense of part (2) of Theorem~\ref{thmSL}.}

\BB{conjecture}{con2}{We conjecture that if $a(r)\in L^1_{loc}$ 
is real, and conditions (I\hskip1pt--V) of the Krein 
theorem hold, 
then $\Pi(\lambda)$ is 
 the limit in average of $p^*(t_n,\lambda)$, that is, 
$$
\Pi(\lambda)=\lim_{r\to\infty}\frac1r\int_0^rp^*(s,\lambda)ds
$$ 
uniformly on compact subsets of~$\mathbb C^+$. Here $a(r)\in L^1_{loc}$ if 
$$\sup_{r\geqslant0}
\int_r^{r+1}|a(s)|ds<\infty.$$ 
If true, this conjecture also implies the uniqueness of $\Pi(\lambda)$. 
Note that in the situation of Theorem~\ref{thm12} the limit 
in average of $p^*(t_n,\lambda)$ does not exists if 
$r_{n+1}- r_n$ are large enough. 
} 
These two conjectures may be related to the results of [{D2}].

%\BB{conjecture}{con11}
%{We conjecture that if $a(r)\in L^p[0,\infty)$ 
% for any $p>2$, and conditions (I\hskip1pt--V) 
%of Krein's theorem hold, 
%then (\ref{eGG}) also holds.} \pagebreak

\end{document}